\documentclass[final,english]{bullsrsl}

\usepackage[utf8]{inputenc}
\usepackage[T1]{fontenc}
\usepackage{natbib}          
\usepackage{graphicx}

\begin{document}
\title{One, two, three, and a tilting frisbee: \\ a transient state of some Be stars}




\author[affil={1}, corresponding]{Dietrich}{Baade}
\author[affil={2,3}]{Jonathan}{Labadie-Bartz}
\author[affil={4}]{Thomas}{Rivinius}

\affiliation[1]{European Organisation for Astronomical Research in the Southern Hemisphere (ESO), Karl-Schwarzschild-Str. 2, 85748 Garching b. München, Germany}
\affiliation[2]{DTU Space, Technical University of Denmark, Elektrovej 327, Kgs. Lyngby, 2800, Denmark}
\affiliation[3]{LIRA, Paris Observatory, CNRS, PSL University, Sorbonne University, Université Paris Cité, CY Cergy Paris University, 5 place Jules Janssen, 92195 Meudon, France}
\affiliation[4]{European Organisation for Astronomical Research in the Southern Hemisphere (ESO), Casilla 19001, Santiago 19, Chile}


\correspondance{dbaade@eso.org}


\maketitle

\begin{abstract}

A recent reanalysis of the extreme variability of $\gamma$\,Cas in the 1930s/40s and 59 Cyg in the 1970s concluded that the disks of these two Be stars have each made a 360-degree turn over the poles \citep{2023A&A...678A..47B}. For a better understanding of the underlying processes and their diagnostic potential for the physics of Be disks, more examples need to be identified, including less spectacular events, in which the disk was just tilted or warped. This paper discusses PZ\,Gem and HD\,120678 as additional stars with spectroscopic evidence of a tilting disk: during an extended period of line emission intermittent shell phases occurred.  In V1294\,Aql, the symptoms of a tilting disk concerned H$\alpha$ only.  Three stars displayed heartbeat-like light variations. Unlike the original heartbeats resulting from distortions of red giant stars in an eccentric orbit, the much-larger-amplitude heartbeats in Be stars are due to changing orientations of the circumstellar disk. Two stars, Pleione and HD\,120678, have exhibited both symptoms so that there is probably just one population of Be stars with tilting disks.  The small number of candidates suggests that large-amplitude tilting or warping of Be disks is not a common process. There is an extreme predominance of spectral types of B1 or earlier. 

In Be binaries resulting from previous mass transfer, spin-orbit misalignments and any torque on the Be disk are not expected to be large. The necessary tilting force could be provided by a putative third star on an orbit highly inclined with respect to the inner binary (or just passing the B star) and be exerted either directly or indirectly via tilting the inner orbital plane.  Over and above a refined knowledge of the multiplicity of Be stars, this would not promise much new insight into Be stars at large (except, perhaps, for some independent estimates of the viscosity parameter and a more 3D view of stellar winds). However, pruning the Be-star saga of such peculiarities may help reduce the level of confusion.   
\end{abstract}



\section{Introduction}
\label{sect:intro}
The picture widely associated with Be stars resembles Saturn: a sphere surrounded by an equatorial disk.  This is Otto Struve's heritage \citep{1931ApJ....73...94S}, who noticed the correlation between the stellar $v$\,sin\,$i$ and the width of the emission lines.   \citet{1989Ap&SS.161...61H} and \cite{2006A&A...460..821A} repeated the analysis with non-photographic data, and, in a higher-level comparison, \citet{2023ApJ...948...34S} found agreement between stellar inclination angles derived from rotational gravity darkening and disk inclination angles obtained from H$\alpha$ emission profiles. 

The existence of such a stellar-circumstellar correlation requires that the disks are generally coplanar with the respective stellar equator.  Interferometry \citep[e.g.,][]{1999A&A...345..203B} and polarimetry (e.g., Reis Querido et al., these proceedings) have confirmed for many stars that the aspect angles of Be disks are $\sim$invariant over many years or even decades.  This is less simple for Saturn, the rings of which can be viewed from below, from above, and even edge-on.  However, the tilting is only apparent and results from the relative orbital motions of Saturn and Earth.  If the disk around a Be star seems to tilt, this explanation is not applicable, and the variation must correspond to a physical change in the angular momentum vector of the disk.  A tilting of Saturn's rings can also be physical in the hypothetical case that another planet passes Saturn closely, in particular if the rings and the motion of the other planet are not coplanar.  

A Be disk (the frisbee in the title) can experience similar encounters. Obviously not in isolated Be stars, but companion stars on misaligned orbits (or a star just passing as in the above planetary example) can inflict transient tilts on Be disks.  If the central B star has received a substantial mass and angular-momentum donation from what now is the secondary star, the spin-orbit misalignment may be small. That is, "one" (see title) is not an option, "two" may be, and "three" certainly is.  The foundation to the suspicion that Be disks may tilt was laid by $\gamma$\,Cas and 59\,Cyg, which in the 1930s/1940s and 1970s, respectively, underwent two transitions each between (i) continuous strong emission and (ii) transient shell phases that are not foreseen in Struve's paradigm and were explained by \citet{1998A&A...330..243H} and \citet{2023A&A...678A..47B} as a temporary tilting of the disks into the line of sight. 

This paper collects and examines public spectroscopic (Sect.\,\ref{sect:spectroscopic}) and photometric (Sect.\,\ref{sect:photometric}) evidence of tilting Be disks.  It discusses the results and summarizes the conclusions in Sect.\,\ref{sect:conclude}.

\section{Be stars with spectroscopic symptoms of a tilting disk}
\label{sect:spectroscopic}
The continuous presence of line emission with just intermittent shell absorption suggests that PZ\,Gem (Sect.\,\ref{sect:PZGem}) 
is cognate with $\gamma$\,Cas and 59\,Cyg.  V1294\,Aql is a much milder case (Sect.\,\ref{sect:V1294Aql}).  Because of the concomitant heartbeat-like photometric variability, HD\,120678 and Pleione, which have gone through similar phase transitions, are included in Sect.\,\ref{sect:photometric}.  Baade, Labadie-Bartz \& Rivinius (in prep.) will present more examples of the spectroscopic category.   

\subsection{PZ Gem (= HD 45314)}
\label{sect:PZGem}
The properties and variability of the O9e star PZ\,Gem were intensively studied by \citet{2015A&A...575A..99R, 2018A&A...615A..44R}.  Optical interferometry \citep{2023A&A...672A...6L} did not find a companion to this X-ray active star \citep{2018A&A...615A..44R}.  From 2014 to at least 2026 \citep{2018A&A...615A..44R, 2026arXiv260909780R} it has a record of uninterrupted line emission (with substantial variations).  In Oct./Nov.\ 2014, there was a brief episode of narrow shell absorption in the Balmer lines and He\,I $\lambda$5876 \citep{2018A&A...615A..44R} so that the outer as well as the inner disk were affected.  Fitting the H$\alpha$ emission profiles, \citet{2018A&A...615A..44R} determined inclination angles between 20$^{\circ}$ and 80$^{\circ}$, in agreement with the argument that transient shell phases in stars with long-term line emission indicate a tilting disk.  They explained their observations with a combination of a warped disk and mass ejections over a range in stellar latitude.  But there is no suggestion for a physical mechanism favoring mass loss at high latitudes.  In contrast, the combination of rapid rotation and nonradial pulsation, which triggers mass loss in some Be stars \citep[][Opli{\v s}tilová et al., these proceedings]{2018pas8.conf...69B}, is most effective near the equator \citep{1997MNRAS.284..839T}.

Around the shell episode, cyclic variations developed in the flux ratio of the violet (V) to red (R) emission components with a rare, huge amplitude, which suddenly set in and equally quickly abated in 2022 \citep{2026arXiv260909780R}.  All standard wisdom about the spectrophotometric variations of Be stars may be invalidated by such extreme V/R activity \citep{Haubois2012}.  However, the variations in $V$-band magnitude and H$\alpha$ equivalent width are not correlated with the V/R ratio.  In spectra from the Be Stars Spectra database \cite[BeSS,][]{BeSS}, the H$\alpha$ emission fluctuated but declined overall, with peak heights of only $\sim$5\% of the ambient continuum level in 2023/2024 but slightly recovering in 2025/2026.  The He\,I and Fe\,II emission lines were more constant in strength than H$\alpha$ so that the dissipation affected the inner disk comparatively little. 

For the explanation of the H$\alpha$ line-profile variations \citep{2018A&A...615A..44R}, \citet{2026arXiv260909780R} considered precession which can also lead to a wide range of emission-line profiles \citep{2022MNRAS.509..931S}.  However, as confirmed by \citeauthor{2022MNRAS.509..931S} with SPH simulations, precession is a slow process, with periods of 20-50$\times$ the orbital period.  For an assumed 100-d orbital period, a precession period of $\geq$5.5\,yr would be expected for PZ\,Gem.  The shell phase may have been as short as 0.2\,yr, which would constrain the angular vertical thickness of the disk to $\leq$15$^{\circ}$.  This is compatible with general observational constraints for Be disks \citep{2015ApJ...799...33C}.  \citet{2026arXiv260909780R} suggested abandoning the precession hypothesis because the observed relative phasing of photometry and line profiles does not match the predictions.  Another possible objection is that the shell episode did not repeat.  For the brightening by 0.3-0.4\,mag in $V$ during the shell phase \citep{2026arXiv260909780R}, there is no conventional explanation.  But a bent disk is a candidate.  

In the time after the shell event, the visual brightness and the equivalent width of the H$\alpha$ emission varied with nearly the same cyclic pattern and were strongly anticorrelated.  According to \citet{1983HvaOB...7...55H}, the latter is the signature of stars viewed edge-on and would imply that the equilibrium orientation of the disk is about edge-on.  This would be a strong argument against disk precession at least after the shell event, and the transient presence of shell absorptions would not require much of a disk tilt if the disk had temporarily grown in geometrical thickness during a mass-loss outburst (see Sects.\,\ref{sect:V1294Aql} and \ref{sect:Pleione}).  The increase in emission strength during the $\sim$100\,d after the shell episode may be a signature of an outburst.  

In summary, the variability of PZ\,Gem is confusing.  The final verdict about the behavior of the disk around the shell event is still out.

\subsection{V1294 Aql (= HD 184279)}
\label{sect:V1294Aql}

For this B0.5 star, \citet{2022A&A...666A.136H} prepared one of the most comprehensive compilations of ground-based photometry and spectroscopy available of any Be star, partly covering up to $\sim$70 years.  
\citet{1983HvaOB...7...55H} found, later confirmed by models \citep{2013ApJ...765...41S}, that Be stars viewed pole-on follow a positive correlation between H$\alpha$ emission strength and $V$-band brightness whereas the correlation is inverse for systems seen edge-on.  Normally, any given Be star belongs to just one of these two groups.  However, \citeauthor{2022A&A...666A.136H} noted \textit{The situation is dramatically different for V1294 Aql}.  \citet{2026A&A...712A.226B} revealed what this was to say: V1294\,Aql alternated between the two patterns.  An unrecognized tilting of the disk could possibly resolve the apparent drama.  This provided an incentive to search for spectroscopic indicators of a varying inclination.  The recently discovered eclipses of the B star by its companion (Labadie-Bartz, Baade \& Rivinius, submitted to A\&A) provides an anchor point for the presumably equatorial equilibrium position of the disk.

The most conspicuous features in the H$\alpha$ emission-strength curves of \citet{2022A&A...666A.136H} are two narrow spikes with a width of a year at their base and reaching $\sim3$$\times$ the previous and subsequent level.  The first event (around RJD\,52,900) was accompanied by significant V/R variability in H$\alpha$ while the second one (around RJD\,55,700) was not.  Accordingly, any underlying variations in the symmetry of the disk produced by $m=2$ disk oscillations \citep{1991PASJ...43...75O, 2008MNRAS.388.1372O} should not have been more than a secondary effect.  Nevertheless, the coincidence of the equivalent-width peak with the central V\,=\,R phase is striking.  

{\bf First event (RJD\,52900):} A short series of BeSS spectra around the maximum showed a transition through a nearly flat-topped H$\alpha$ profile with peak flux an order of magnitude higher than that of the continuum, which is exceedingly rare among Be stars.  Such profiles can develop if the disk inclination is $\sim$45$^{\circ}$ or even less \citep{2020ApJ...894...18S, 2010ApJS..187..228S}.  In all other BeSS spectra (from mid-2003 to mid-2026, with large gaps between them), there was a central depression in H$\alpha$, which reached large depth at times, but never got close to the ambient continuum level and so did not satisfy the shell criterion proposed by \citet{1996A&A...308..170H}.  The available profiles of HeI $\lambda$6678 always had a shell appearance.  FeII and non-H$\alpha$ Balmer lines were only observed 15 months before and  1-2 months after the peak of the event.  Their shell nature was well developed, as was the case at all other times, often flanked by emission components with varying V/R ratios.  

{\bf Second event (RJD\,55,700):}  The BeSS spectra of H$\alpha$ and HeI $\lambda$6678 closest in time are from 9 months before and 43 months after it.  The depression in H$\alpha$ was deep in both cases but well above the continuum level.  The persistent shell character of the Balmer lines other than H$\alpha$ was already mentioned above.  In all Ond{\v r}ejov and La Silla spectra used by \citet{2022A&A...666A.136H}, HeI $\lambda$6678 had a shell profile except for the onset of the event (on RJD\,55,688 and 55,692), when, in addition, the two similar emission peaks reached their highest level.  Prior to that, on RJD\,55,398, the shell nature was still there, by RJD\,55,740 it had returned. On these four dates and through $\sim$RJD\,60,000, H$\alpha$ was in pure emission (no shell).


The photometric coverage of the two events is incomplete \citep{2022A&A...666A.136H}, but they have a strong increase (reddening) in \textit{U-B} by more than 0.5\,mag in common.  In the first event, \textit{B-V} became bluer by 0.1\,mag, and the star faded in $V$ by about 0.4\,mag.  For the second event, one can only guess that the variations in \textit{B-V} and $V$ were roughly the same as in the first one.  


In summary, a common pattern emerges \citep[see also][]{2026A&A...712A.226B, 1982BAICz..33..308H}: The star dims, gets slightly bluer in \textit{B-V}, reddens strongly in \textit{U-B}, and the Balmer emission increases.  Except for H$\alpha$, shell components are permanent during and outside the events, and double-peaked emission components are frequent.  
%
%
Overall, this is consistent with the inverse correlation described by \citet{1983HvaOB...7...55H} for H$\alpha$.  But in V1294\,Aql, notably H$\alpha$ itself is exempt from the agreement.  Again except for H$\alpha$, the pattern is also similar to 88\,Her (Sect.\,\ref{sect:Pleione}) and, in contrast to the recorded talk associated with this paper, suggests that all three shell cycles were the responses of the disk to phases of enhanced mass loss (outbursts).  During these events, the inner disk probably puffed up and intersected the line of sight, strengthening the shell lines. 

\citet{1983HvaOB...7...55H} presented V1294\,Aql as an example of his anticorrelation pattern, but this star's H$\alpha$ profile has exhibited non-conforming peculiarities since.  In the first event (the second was not covered by BeSS), it developed what almost looked like a huge single peak signaling a much reduced disk inclination.  But all other lines had shell profiles.  
Only between RJD\,60,000 and 61,000, when the disk was decaying, did H$\alpha$ finally develop a clear shell core. 

Overall, the inner disk, i.e., the disk without the part contributing the bulk of the H$\alpha$ emission, did not undergo significant variations in its position relative to the line of sight.  This is supported by the absence of changes in emission width. 
The observations of the H$\alpha$ profile are in strong conflict with this stationarity and suggest that the inclination of the outer disk changed by several tens of degrees.  The persistent edge-on view of the inner disk provides for a robust absolute comparison (Baade, Rivinius \& Labadie-Bartz, in prep.). 

Although symptoms of a tilting/warping disk were found indeed, it remains unclear on what basis \citet{2022A&A...666A.136H} and \cite{2026A&A...712A.226B} concluded that V1294\,Aql exhibited an inverse as well as a positive correlation between $V$-band brightness and H$\alpha$ emission strength.  During the two outbursts, it was inverse.  In the past decade, the disk has been dissipating, and the data collected by \citeauthor{2022A&A...666A.136H} document an inverse correlation for this phase, too.  Conceivably, \citeauthor{2026A&A...712A.226B} had the color behavior in mind.  In the \textit{U-B} vs.\ \textit{B-V} diagram, the star migrated through the regions occupied by edge-on and by pole-on disks.  It should be reminded that after Harmanec found his correlations, it was learned that the H$\alpha$ emission forms in the outermost disk \citep[e.g.,][]{1997ApJ...479..477Q} while the \textit{UBV} flux from the disk arises from the innermost part \citep{Haubois2012}.  Today, one would not couple these parameters because time delays between their regions of formation in their response to structural changes can dilute any correlation, and if the disk tilts or warps, the physical basis can get lost altogether.  

In the title to their paper, \citeauthor{2022A&A...666A.136H} speculated that V1294\,Aql might be \textit{an important clue to the Be phenomenon}, and in the text announced: \textit{Attempts at modelling} [the variations] \textit{quantitatively, planned for a continuation of this study, are expected to shed more light on the mysterious Be phenomenon.}  This is not likely because $\sim$10 days after its injection into the disk, matter no longer remembers where and how this happened \citep[cf.\,][]{2025A&A...699A..82L, 2026arXiv260705270R}.  The photometry used by \citeauthor{2022A&A...666A.136H} does not trace variability on this timescale.  In some Be stars, rapid multi-periodic variability only observable from space is getting close to explaining the Be phenomenon as the result of the temporary superposition of multiple nonradial-pulsation modes \citep[][Opli{\v s}tilová et al., these proceedings]{2018pas8.conf...69B}.

\section{Be stars with photometric symptoms of a tilting disk: the heartbeat disks}
\label{sect:photometric}
By analogy to heartbeat stars (Sect.\,\ref{sect:heartbeat}), this section introduces the concept of heartbeat disks.  Heartbeat disks can mimic the LCs of heartbeat stars but with the tenfold larger amplitudes observed in Be stars.  This is so because (i) a star passing the Be star can inflict much larger distortions on its thin and extended low-mass circumstellar disk than on the central near-main-sequence B star and (ii) a disk with changing inclination angle redirects significant, variable fractions of the stellar flux to the observer.  {The large heartbeat amplitudes observed in Be stars also safely distinguish them from the small pulsational variability \citep[][Opli{\v s}tilová et al., these proceedings]{2022AJ....163..226L}.

\subsection{Heartbeat stars}
\label{sect:heartbeat}
Heartbeat stars were named by \citet{2012ApJ...753...86T}, who perceived a similarity of their LCs to heartbeats in human electrocardiograms.  Meanwhile, their number exceeds 1000 \citep{2022ApJS..259...16W} with very diverse LCs: single-peaked, double-peaked, and structures reminiscent of regular and inverse P Cygni spectral line profiles.  The variability is predominantly due to the changing shapes of stars, which are gravitationally distorted in eccentric binaries \citep{2017MNRAS.472.1538F}.  Since this works more effectively for stars with low surface gravity, 90\% of them are red giants, and the others are in the Hertzsprung gap.  The shape of the LC depends sensitively on the orientation of the orbit with respect to the observer so that modeling can derive the eccentricity, the inclination, and the argument of periastron \citep{2020ApJ...903..122C}.  Typical amplitudes are a few 0.01\,mag, and most known periods fall in the range from a week to a year.  The close orbital passages can also tidally excite oscillations (TEOs) with frequencies that are integers of the orbital frequency and, in most cases, with sub-mmag amplitudes.

\subsection{HD 120678}
\label{sect:HD120678}

This Galactic O9e star is one of the two Rosetta stars of the sample (the other one is Pleione, see Sect.\,\ref{sect:Pleione}) as it combines a heartbeat LC with a transient shell phase embedded in an extended period of persistent line emission.  In 2008, \citet{2012A&A...546A..92G} discovered shell-like absorption components, which replaced many of the Oe-typical emission lines and appeared as additional features in Balmer lines.  At 200\,km\,s$^{-1}$, the new absorption lines were about half as wide as the photospheric HeII absorptions and, therefore, much broader than typical Be-shell lines \citep{2006A&A...459..137R}.  If the geometry of their formation is analogous to that of permanent Be-shell stars, the larger breadth would indicate a region close to the star (unless the volume over which the features formed was dynamically perturbed).  The observers found the excitation level of the shell-like metal lines intermediate between those of the star and the circumstellar disk.  The spectra closest in time were obtained one month before and one year after the event.   Because of their lower resolution, weaker shell components might have gone unnoticed.  However, the combination of their transient nature (the shell-like event) with the permanence of the line emission is the tell-tale signature of a tilting (or warping) disk.  

What makes the event so remarkable is the LC.  It resembled an inverse P Cygni profile, which is one of the standard shapes of stellar heartbeat LCs \citep{2022ApJS..259...16W}.  It spanned $\sim$4 years and left the brightness outside the event roughly unchanged.  The minimum coincided to within a day with the shell-like event.  This can be expected because, when a star is viewed through an unbent edge-on disk, the extinction is largest, and the visible disk area emitting reprocessed stellar flux is smallest.  

The assumption of a tilting disk predicts that emission profiles observed close to the shell phase are wider than during quiescence.  In HD\,120678, this would concern the four spectra reproduced by \citeauthor{2012A&A...546A..92G} for the time from a year to little more before the appearance of the anomaly.  Unfortunately, the considerable V/R variability would render any width estimates very uncertain.  In spite of their lower resolution, the Balmer profiles observed just a month before the shell were without doubt intrinsically strongly broadened.  Accordingly, both photometry and spectroscopy imply that the disk was already substantially tilted prior to the shell-like event, which followed when the Balmer-emitting part of the disk was tilted into the sight line.  During the event, the H$\beta$ emission was equally wide, but the peak heights were reduced to one-third, matching a typical Be-shell appearance.  

It is unknown whether HD\,120678 is a binary.  \citet{2012A&A...546A..92G} derived a tentative period of 280\,d but did not publish the radial-velocity curve.  A paper dedicated to the tilting disk around HD\,120678 just appeared \citep{2026A&A...715L...8B}.

\subsection{Pleione}
\label{sect:Pleione}
The photometric, spectroscopic, and polarimetric variability of this late B-type SB1 star with a permanent disk and transient shell phases has marred Be-star research for decades.  \citet{2022ApJ...928..145M} seem to have recently solved key elements of the puzzle.  In their model, the disk is in a 34-yr oscillatory fashion tilted by a companion on an eccentric and misaligned 218-d orbit. The length of the cycle is governed by the time it takes for the secondary to tear off the outer part of the disk, which then begins to precess freely about the B star and ultimately dissolves.  The inner disk returns to its equatorial equilibrium position, and the tilting cycle starts anew.  The model explains (most of) the photometric variability, which has a peak-to-peak $V$-band amplitude of $\sim$0.4\,mag, through the changing aspect angle of the disk. 

For the tearing of the disk, \citet{2022ApJ...928..145M} refer to SPH simulations by \citet{2022MNRAS.509..931S}, which show that after dozens of orbits, a companion on an orbit strongly misaligned with the primary's equatorial plane, can tear a disk.  If the 34-yr cycle length is a period, this explanation should not work for early-type Be stars because their variable star-to-disk mass transfer typically lets the disks strongly grow and shrink on timescale of years.  Late-type Be stars like Pleione appear much less active \citep{2013A&ARv..21...69R}, and Pleione has shown little irregularity in its H$\alpha$ equivalent width \citep{2022ApJ...928..145M}.  However, should such irregularity develop without affecting the 34-yr cycle length, a third star could offer the clocking.  

In the on-line recording accompanying this paper, it is said that the variability of 88\,Her is very similar to that of Pleione.  This concerns the shape of the $V$-band LC and the coincidence of fadings in $V$ with strengthenings of shell lines.  However, these cyclic variations are probably due to well-isolated outbursts \citep[][see also Sect.\,\ref{sect:V1294Aql}]{Haubois2012}, and the main empirical photometric property distinguishing the two stars lies in their color curves \citep{2026A&A...712A.226B}.

\subsection{AzV 493}
\label{sect:AzV493}
This O4-5e star in the SMC is one of the hottest ($T_{\rm eff}$\,$\approx$\,42,000\,K) high-luminosity stars known \citep[][see also Oey et al., these proceedings]{2023ApJ...947...27O, 2026PASP..138g4202O}.  It has displayed a heartbeat LC with amplitude $>$1\,mag and inverse P Cygni shape in the $I$ band.   The first part of the second cycle, in which the "P Cygni emission" peak of a periodic LC would have appeared, was wiped out by oscillatory other variability.  Accordingly, the period (if any) could be 7.3 or 14.6\,yr.  The most recent LC shows the beginning of a new cycle, which does repeat the initial heartbeat-like brightening seen in the first cycle.  Shell lines have not been reported.  If the star has a Be-star-typical low-mass companion as a former mass and angular-momentum donor, and the possible 7.3/14.6\,yr period is due to an outer companion, the system is at least triple.  Radial-velocity measurements and a single X-ray observation near what might be the periastron did not clarify the multiplicity status of the star.  The faster (30-45\,d) and lower-amplitude ($\sim$0.2\,mag peak-to-peak) oscillatory photometric variability appears to be unrelated to the disk heartbeat and is shared with several other Be stars.  Stars with symptoms of a tilting/warping disk among them are $\gamma$\,Cas \citep[][55-73\,d]{2002ApJ...575..435R}, HD\,120678 \citep[][$\sim$300\,d; see also Baade, Labadie-Bartz \& Rivinius, subm.]{2004AcA....54..153P}, and PZ\,Gem \citep[][230\,d]{2026arXiv260909780R}.  The example of AzV\,493 demonstrates that this variability is not always visible.  For the lack of a reference survey, it cannot be said whether the implied fraction is high.

\subsection{MACHO 80.7443.1718}
\label{sect:MACHO}
This single-lined B0\,Iae binary with a period of 32.8\,d in the LMC has been called the "loudest heartbeat star" \citep{2021MNRAS.506.4083J}.  On account of the peak-to-peak amplitude of $\sim$0.5\,mag, this may be justified because the typical amplitudes of heartbeat stars are only a few 0.01\,mag (but an apparent order-of-magnitude jump in a standard quantity may also produce noise outside the data).  The light-curve shape is of the "normal P Cygni" type.  Although some of their team members \citep{2022A&A...659A..47K} had detected TEOs in this star, confirming the strong tidal impact, \citet{2024A&A...686A.199K} rejected the heartbeat interpretation because it is not in agreement with the achromaticity of the main variability.  Other modelers took the heartbeat hypothesis at face value and inferred monstrous breaking waves at the surface of the primary \citep{2023NatAs...7.1218M, 2026ApJ..1006..158M}.  

Perhaps the decisive observation for the understanding of this unusual star is the Balmer line emission.  Its double-peak structure is reminiscent of classical Be stars, and the variation of the ratio in strength of the two components is qualitatively identical to V/R variation in Be stars.  However, circumstellar disks are not common in B-type supergiants \citep{1981ApJ...247..158S, 1983A&A...124..211B}.  In this star, a disk may be due to decretion facilitated by the extremely rapid rotation ($v$\,sin\,$i$ $\approx$ 175\,km\,s$^{-1}$) of the primary and further boosted by the companion in a tight orbit.  The V/R variability has resisted modeling efforts \citep{2024A&A...686A.199K}.  These did not consider the possibility of a tilted disk and did not simultaneously target the photometric variability.  However, the case of Pleione (Sect.\,\ref{sect:Pleione}) suggests this to be tried.

\section{Discussion and conclusions}
\label{sect:conclude}

Tilting disks around Be stars are an infrequent but not exotic phenomenon.  There are two identifying symptoms, photometric heartbeat LCs and the transient appearance of shell absorptions in spectra.  Heartbeats are a newly recognized type of photometric variability of Be stars.  In HD\,120678 \citep{2026A&A...715L...8B} and Pleione, both a photometric heartbeat and transient shell absorption have been observed.  Therefore, the two types of variability should not be attributed to two types of star but belong to one population, which is not systematically different from other classical Be stars.  Their combined appearance in HD\,120678 and Pleione leaves little doubt that their disks tilted as coherent structures.  Whether both symptoms or only a heartbeat LC can be observed depends on if the disk crosses the line of sight.  

The disk's contribution to the $V$-band flux originates from its innermost $\sim$ two stellar radii \citep{Haubois2012} while the H$\alpha$-emitting part can be an order of magnitude larger.  Therefore, a bending disk can produce seemingly conflicting symptoms.  

The argument of intermittent shell phases as indicators of tilting disks must be used with care.  The example of 88\,Her illustrates that shell lines may be transient in stars viewed edge-on but do not imply 3D motions of the (full) disk plane.  In these stars, shell phases may be associated merely with enhanced mass loss from the star.  This may temporarily increase the thickness of the disk and the probability of the sight line to the star passing through it. V1294\,Aql combines this picture with a warping of the H$\alpha$-emitting, outer region.  

None of the Be stars considered above are known to be a triple system, but a second companion can explain tilts of disks quite easily while post-mass-transfer binaries probably require numerical simulations.  $\nu$\,Gem \citep{2021ApJ...916...24K} and $\omicron$\,Cas \citep{2026A&A...709A.167H} are confirmed triple stars with a Be component, and $\omicron$\,And is even in a quadruple system \citep{1988PASP..100..243H, 2010ARep...54.1134Z}.  In all these systems, the Be star is in a binary of a wide hierarchical system.  If these Be stars have formed through mass transfer from what today is an sdO/B star and this low-mass star has not been ejected, the stated multiplicity level would increase by one.  Gaia DR4 will soon provide more insights \cite[see also][]{2024MNRAS.527.3076D}.  

With the exception of Pleione, all other classical Be stars named above and known or strongly suspected to have experienced a tilting of their disk are of very early spectral type (AzV\,493: O4-5, PZ\,Gem: O9, HD\,120678: O9, $\gamma$\,Cas: B0.5, V1294\,Aql: B0.5, 59\,Cyg: B1), at least as early as B1, where the frequency of Be stars relative to disk-less B stars of luminosity classes III-V peaks in the solar neighborhood \citep[][most of the original data are from the photographic era]{1997A&A...318..443Z}. This frequency drops steeply towards higher temperatures.  The reason is not established, but the larger abundance of Oe stars in the Magellanic Clouds suggests metallicity as a contributing factor (weaker winds?).  Another difference setting Pleione apart is that its shell phases are predictable and the companion plays a big role in the disk cycle.  However, if the cycle length is stable at 34\,yr, a third star may be a convenient hypothesis for the explanation.  In all the other stars with shell phases, these events never repeated and were so isolated in time that a critical involvement of a single companion with a Be+sdO/sdB-star-typical (or Be+WD) binary period is hard to imagine. Postulating a trigger by a passing third star would combine arbitrariness with physical simplicity.  

In Pleione, the disk inclination changes over decades with every revolution of the companion on a misaligned orbit.  Misalignment is not likely to explain the much faster inner part of the one-off heartbeat LC of HD\,120678.  A large eccentricity, possibly even $>$\,1, and in any event outside the range of typical Be+sdO/B or Be+WD  binaries, may have to be considered.  

The dominance of very hot Be stars may be just a statistical fluke.  There may also be an observational bias because O stars are intrinsically brighter.  But one may also take into account that the multiplicity of massive stars is high \citep{2017ApJS..230...15M} and that disks around massive Be stars may be larger and, therefore, more susceptible to external perturbations.  

Tilting disks do not seem to be relevant for the understanding of the Be phenomenon (the development of gaseous disks around B stars).  But their observations and modeling may enable measurements of the viscosity in response to forces outside the equatorial plane.  It is also conceivable that a disk turning out of the line of sight may reveal the stellar wind in the equatorial plane so that the latitudinal dependence of the wind can be studied.  This will require simultaneous quantitative modeling and disentangling of variable wind and disk lines.

This paper gives preference to Be stars with photometric symptoms of a tilted (heartbeat) disk while there are more stars, for which only spectroscopic symptoms are known.  A paper describing them is in preparation (Baade, Labadie-Bartz \& Rivinius). 





\begin{acknowledgments}
We thank the anonymous reviewer for useful questions.  This work has made use of the BeSS database (operated at LESIA, Observatoire de Meudon, France, http://basebe.obspm.fr), the SIMBAD database and the VizieR catalogue access tool (offered by CDS, Strasbourg Astronomical Observatory, France DOI : 10.26093/cds/vizier), and the Astrophysics Data System, funded by NASA under Cooperative Agreement 80NSSC21M0056.  
This research was funded/co-funded by the European Union (ERC, MAGNIFY, Project 101126182). Views and opinions expressed are however those of the author(s) only and do not necessarily reflect those of the European Union or the European Research Council. Neither the European Union nor the granting authority can be held responsible for them.
\end{acknowledgments}

\newpage
\begin{furtherinformation}

\begin{orcids}

\orcid{0000-0003-1637-9679}{Dietrich}{Baade}
\orcid{0000-0002-2919-6786}{Jonathan}{Labadie-Bartz}
\orcid{0000-0003-1013-5243}{Thomas}{Rivinius}
\end{orcids}

\begin{authorcontributions}
DB: conceptualization, investigation, writing - original draft; 
JLB and TR: conceptualization, investigation, writing - review and editing.  
\end{authorcontributions}

\begin{conflictsofinterest}
The authors declare that there is no conflict of interest.
\end{conflictsofinterest}

\end{furtherinformation}



%

\bibliographystyle{bullsrsl-en}

\bibliography{DBAADE.bib}

\end{document}